\documentstyle[epsfig]{europhys}

\def\And{{\rm and\ }}

\def\stars{\bigskip\centerline{***}\medskip}

\newif\ifboo \boofalse

\def\Name#1{{\sc #1},}
\def\Vol#1{\ifboo Vol. {\bf #1}\else{\bf #1}\fi}
\def\Year#1{\ifboo #1\else(#1)\fi}

\def\Page#1{\ifboo {\rm p. #1}\else{\rm #1}\fi}

\begin{document}
\euro{0}{0}{0-0}{1999}
\Date{July $4^{th}$ 2000}
\shorttitle{}
\title{Enhancement of structural rearrangement in glassy systems 
under shear flow}
\author{F.Corberi\inst{1}, G. Gonnella\inst{2}
  \And D.Suppa\inst{2} }
\institute{
     \inst{1} Istituto Nazionale per la Fisica della Materia,
Unit\`a di Salerno {\rm and} Dipartimento di Fisica, Universit\`a di Salerno,
84081 Baronissi (Salerno), Italy \\
     \inst{2} Istituto Nazionale per la Fisica della Materia,
     Unit\`a di Bari {\rm and}
Dipartimento di Fisica, Universit\`a di Bari 
{\rm and} Istituto Nazionale di Fisica Nucleare, 
Sezione di Bari, \\ 
via Amendola 173, 70126 Bari, Italy }
\rec{14 December 1999}{in final form 12 September 2000}
\pacs{
\Pacs{64}{70.Pf}{Glass transitions}
\Pacs{05}{70.Ln}{Non-equilibrium thermodynamics, irreversible processes}
\Pacs{83}{50.Ax}{Steady shear flows}
      }
\maketitle
\begin{abstract}
We extend the analysis of the mean field schematic model 
recently introduced \cite{1} for the description of 
glass forming liquids to the case of a supercooled fluid 
subjected to a shear flow
of rate $\gamma$.
After quenching the system to a low temperature $T$, 
a slow glassy regime is observed before
stationarity is achieved at the characteristic 
time $\tau _g$.
$\tau _g$ is of the order of the 
usual equilibration time without shear $\tau _g^o$ 
for weak shear,
$\gamma \tau _g ^o<1$. For larger 
shear, $\gamma \tau _g ^o>1$,
local rearrangement of dense regions is instead enhanced 
by the flow, and $\tau_g \simeq 1/(T\gamma) $.

\end{abstract}

Structural rearrangement in supercooled liquids and glassy systems is
severely suppressed due to configurational restrictions, requiring
a cooperative dynamics of correlated regions that involves many degrees
of freedom \cite{2}. 
This complex behaviour often results in a slow 
kinetics characterized by diverging relaxation times at 
the temperature of structural arrest $T_o$ \cite{3} 
and strong non-equilibrium effects,
such as aging. 
A glassy system above $T_o$ 
is generally observed to be off-equilibrium either because
a modification of the control parameters, such as the pressure or the 
temperature, has been exerted or because it is driven mechanically.
In the latter situation it is possible to show 
that the aging of some systems can be 
triggered by the external forcing and that they look 
younger if a larger 
drift is applied \cite{4}. 
This is also witnessed by the modalities of the violation of the fluctuation 
dissipation theorem \cite{5,6}. In many cases, by injecting power 
there is the possibility to interrupt aging and stabilize
the system into a power-dependent stationary state.
This effect is observed, for instance, in gelling systems
under an applied shear force \cite{gel}.

In this letter we study the out of equilibrium evolution of a  
glassy system, such as a supercooled fluid,  quenched above $T_o$ 
in the presence of a shear flow with rate $\gamma$ \cite{5b}.
The analysis of the dynamics is carried out in the framework of a 
model \cite{1} recently introduced for the description
of the glassy behavior close to the dynamical transition. 
The approach is simple
enough to be handled analytically, allowing explicit calculations.
Our main result is the observation of two regimes:
a weak shear regime, where the flow
does not practically affects the glassy behavior, and a 
strong shear situation, where the inner relaxation process is enhanced, because
density fluctuations are convected by the flow, and aging is interrupted
after a time  $\tau_g \simeq 1/(T\gamma) $.
For $t>\tau _g$ a shear-induced time-translational invariant steady state is
entered where the decay of the two-time correlation
function is characterized by the presence of
inflection points. 

When a macroscopic motion of the fluid is present, the mean 
field version of the model introduced in \cite{1} is generalized
by the following constitutive Equation
\begin{equation}
\frac {\partial \rho (\vec r,t)}{\partial t}+\vec \nabla \cdot 
[\rho (\vec r,t)\vec V(\vec r)]=
D(t)\nabla ^2  \rho (\vec r,t) +\eta (\vec r,t) 
\label{meanf}
\end{equation}
where  $\rho (\vec r,t)$ is a coarse grained particle density,
and $D(t)=\langle M(\rho)\rangle$ is the 
average mobility of the particles. $\eta$ is a 
gaussianly distributed random field, representing thermal noise, with
expectations $\langle \eta (\vec r,t)\rangle=0$ and
$\langle\eta (\vec r,t)\eta(\vec r',t')\rangle=-2TD(t)\nabla ^2
\left [ \delta (\vec r-\vec r')\delta (t-t')\right ]$, where 
$\langle ...\rangle$ is the ensemble average and $T$ is the
temperature of the bath.
In Eq.~(\ref{meanf}) the second term on the l.h.s. is an advection
contribute \cite{7}, due to the presence of the flow, which was not
considered in \cite{1};
$\vec V$ is the velocity of the fluid: for the
case of a plane shear flow we consider $\vec V(\vec r)=\gamma
y\vec e_x$, $\vec e_x$ being the unitary vector in the flow direction.
Eq.~(\ref{meanf}) describes a system where the shear rate is
homogeneous throughout the sample and applies therefore only to materials
which can support such a flow (these are sometimes 
referred to as soft
glassy materials, as opposed to hard glassy materials where strain localization
or fractures are observed \cite{8}). 

The present model aims to describe the main features of the out-of-equilibrium
dynamics above $T_o$. It is schematic in spirit and, in order to be 
generic, leaves aside as much system specific details as possible. 
The basic assumptions is that a good deal of the
complex behavior of glassy systems can be encoded into 
the conventional convection-diffusion \cite{7} equation~(\ref{meanf})
by means of a suitably chosen particle mobility $M(\rho)$.
In \cite{1} a 
quickly vanishing function of the density, $M(\rho )=\exp \{ v[\rho-1]^{-1}\}$
was proposed on phenomenological grounds;
here $v$ is a (temperature dependent) parameter and the particle density
has been rescaled so that $\rho =1$ is the point of dynamical arrest.
This form of the mobility has been obtained in different approximations
by several authors in apparently heterogeneous contexts 
as the free-volume theory 
of the glass transition \cite{2} 
or in "car parking" problems in one dimension \cite{9}.
In the mean field version of the model, expressed by Eq.~(\ref{meanf}), 
$M(\rho)$ is replaced by its
average value that can be computed as
\begin{equation}
D(t)=[2\pi S^2(t)]^{-1/2}\int _0 ^{1}
M(\rho)e^{-(\rho -\overline \rho)^2/[2S^2(t)]}d\rho
\label{effdiff}
\end{equation}
where  $\overline \rho =\langle \rho \rangle$ and
$S^2(t)=\langle (\rho -\overline \rho ) ^2 \rangle$,
since the density distribution is gaussian \cite{1}. 

From Eq.~(\ref{meanf}), by transforming into momentum space,
the following formal solution for the two-time correlator 
${\cal C}(\vec k',\vec k, t_1,t_1+\Delta t)=
\langle \rho (\vec k',t_1)\rho (\vec k,t_1+\Delta t)\rangle
= C(\vec k,t_1,t_1+\Delta t) \delta \left [ \vec {\cal K} (\Delta t)
+\vec k' \right ]$ is found, with
\begin{equation}
C(\vec k, t_1,t_1+\Delta t)= 
\chi [\vec {\cal K}(\Delta t),t_1]
\exp \left [-\int _0 ^{\Delta t} 
{\cal K}^2 (s)D(t_1+\Delta t-s)ds \right ]   
\label{meanfc}
\end{equation}
where $\vec {\cal K}(s)=\vec k +\gamma k_xs\vec e_y$, 
$\vec e_y$ being the unitary vector in the shear direction. 
Notice the presence of the delta function  
$\delta \left [ \vec {\cal K} (\Delta t)+\vec k' \right ]$, as opposed
to the usual $\delta (\vec k+\vec k')$, due to the distortion
induced by the flow \cite{10}.
$\chi (\vec k,t)=C(\vec k, -\vec k,t,t)$ 
is the structure factor that,
with a high temperature disordered initial condition $\chi (\vec k,0)=\Delta$,
evolves according to
\begin{equation}
\chi (\vec k,t)=(\Delta -T)e^{-2\int _0 ^t {\cal K}^2 (s)D(t-s)ds}+T
\label{formalchi}
\end{equation}
From Eq.~(\ref{formalchi}) the average density fluctuations can be computed 
through \\
$S^2(t)=(2\pi)^{-d}\int _{|k|<\Lambda} \chi (\vec k,t) d\vec{k}$, 
where $\Lambda$ is a phenomenological momentum cutoff.
For sufficiently long times the integral of the exponential terms in
Eq.~(\ref{formalchi}) can be extended to the whole $\vec k$-plane
(letting $\Lambda=\infty$) because the support of $\chi$ shrinks towards
the origin (see Fig.~1), yielding
\begin{equation}
S^2(t)= \frac{\pi ^{-\frac {d}{2}}}{2^d}(\Delta-T)R_0(t)
^{\frac{1-d}{2}} \cdot 
\left \{R_0(t)-\gamma ^2\left [ \frac {R_1^2(t)}{R_0(t)}-R_2(t) 
\right ]\right \}  ^{-\frac{1}{2}} +qT
\label{esse}
\end{equation}
where $q=(\Sigma _d /d)[\Lambda /(2\pi)]^d$,
$\Sigma _d $ is the surface of the $d$-dimensional unitary hypersphere
and $R_n(t)=2\int _0 ^t z^n D(t-z)dz$.
Eqs.~(\ref{effdiff},\ref{esse}) are closed coupled equations which allow
the computation of $D(t)$ and hence the whole evolution of the model, 
through the correlator (\ref{meanfc}).

In the case with no flow it has been shown \cite{1} 
that when $\overline \rho$
is close to the critical value $\overline \rho =1$ 
a {\it cage effect} produces a transient pinning phenomenon
characterized by the constancy of the main observables, namely
$C,\chi,D$ and $S$.
Then, for longer times, less dense regions start evolving, but yet 
high density regions
are almost frozen and evolve slower, 
producing a glassy behavior characterized by
the existence of many time scales and aging. 
The evolution of the system in this regime is characterized by a 
decay of $C$ slower than the
usual diffusive behavior with
an exponential damping of the correlations.
Aging is eventually interrupted
after a characteristic time $\tau _g^o$ and an equilibrium state is entered
where time translational invariance is recovered. For dense
systems \cite{11}, equilibration is induced by thermal fluctuations
which shake the frozen regions of the system.

When shear is applied to the fluid a similar situation occurs.
In the following we describe the main results of the analytical 
solution.
We refer to a deep quench (low $T$) with a high density
($\overline \rho \simeq 1$). 
Initially the {\it cage effect} is observed, as discussed in \cite{1};
this is reflected in Fig.~2 by the constancy of $D$ for small times. 
For longer times the evolution starts and the glassy regime is entered,
as for $\gamma =0$. The behavior of the structure factor in this time domain
is shown in Fig.1 where the anisotropic character of the correlations 
is evident.
We consider the situations relative to different 
ranges of the strain $\gamma t$.

{\bf Small strain.}
For sufficiently small $\gamma t$, neglecting
the terms proportional to $T$ and $\gamma $ in Eq.~(\ref{esse}) 
one can solve Eqs.~(\ref{effdiff},\ref{esse}) obtaining 
\begin{equation}
D(t)\sim t^{-1}(\ln t)^{\delta -1},
\end{equation}
where $\delta=6/d$, as without flow.
This is shown in Fig.~2 where it is seen that 
the curves for $D(t)$ initially
collapse onto the $\gamma =0$ line.
Inserting the expression for $D$ into  Eq.~(\ref{meanfc}) 
one finds that the terms proportional to 
$\gamma $ are negligible and $C$ falls isotropically as an enhanced power law.
Time translational invariance is lacking and the system ages.
Shear induced effects become relevant at the crossover time $\tau _c$ when 
the terms proportional to $\gamma $ and $T$ cannot be neglected in
Eq.~(\ref{esse}).
From the analysis of Eq.~(\ref{esse}) we obtain
 
\begin{equation}
\tau _c =\min \{ \tau_g^o , \gamma ^{-1} \}
\label{tau}
\end{equation}
and $\tau_g^o$, the time at which
aging is interrupted in the corresponding undriven system, is computed
in \cite{1}.
In the weak shear regime, with $\tau_g^o <<\gamma ^{-1}$, 
the glassy evolution is ended
at $t\simeq \tau_g^o$, due to the thermal fluctuations, and the presence of
the flow only affects the asymptotic dynamics that will be discussed below.
For the choice of parameters of Figs.~1,2 one has $\tau_g^0\simeq 10^8$ 
(this is the time at which $D$ approaches the asymptotic constant
value) and so, from Eq.~(\ref{tau}), 
we expect the curves of $D$ to depart from the $\gamma =0$ case for 
$\gamma \leq 10^{-8}$, as shown in Fig.~2.

{\bf Large strain.}
For large strain and values of $\gamma$ in a strong shear regime with 
$\gamma ^{-1} << \tau_g^o$, 
the glassy behavior is changed  for $t> \tau _c$ by the motion of
the fluid.
By keeping only terms proportional to $\gamma$ in Eq.~(\ref{esse}) we find
\begin{equation}
D\sim e^{-(t/t_o)^{1/3}}, 
\end{equation}
where $t_o$ is a constant.
In this glassy regime the convergence to the
asymptotic stationary state is enhanced by the shear. 
One can check that the term
$qT$ is negligible on the r.h.s. of Eq.~(\ref{esse}) up to 
$\tau_g=(T\gamma)^{-1} $,
when aging is interrupted. 
The dependence of $\tau_g$ on $\gamma$ is shown in the inset of
Fig.~2, showing that $\tau_g =\tau_g^0$ for small $\gamma $ while it 
approaches a $\tau_g \sim \gamma^{-1}$ law for larger values of the shear.
This completes the description of 
the pre-asymptotic glassy stage.

We consider now the behavior of the system in the asymptotic domain 
($t>\tau _g$). 
Since $R_n(t)\to \infty$, one has $S(t)\simeq qT$
so that $D(t)=D(\infty)=const.$. This can be seen in Fig.~2.
Hence one time quantities do not
depend on $t$ and two time observables, such as the correlator~(\ref{meanfc}),
are functions of the difference $\Delta t$ alone
\begin{equation}
C(\vec k,\Delta t)=T\exp \{ -D(\infty)[2\Delta t k^2+
\gamma (\Delta t)^2 k_x k_y +
\frac{\gamma ^2}{3}(\Delta t)^3 k^2_x ]\} 
\label{trans}
\end{equation}
Time translational invariance is then recovered. For $\gamma =0$,
from Eq.~(\ref{trans}), $C$ has a simple exponential form.
When the shear is present, on the other hand,
the decay is faster for large $\Delta t$ (except at $k_x=0$), indicating
that the relaxation of fluctuations is enhanced by the flow. Actually
the behavior of $C$ is more complex, as shown in Fig.~3.
In this picture the behaviour of $C(\vec k,\Delta t)$
is plotted at  $\gamma=10^{-2}$ for different $\vec k$.
In the sectors $k_xk_y>0$ all the coefficients of the polynomial form
in the argument of the exponential in Eq.~(\ref{trans}) are positive.
Then $C$ is depressed with respect to
the case $\gamma =0$ for any $\Delta t$.
On the other hand, $C$ develops  
inflections in the sectors with $k_xk_y<0$.
From the comparison with the case $\gamma =0$ one sees
that this effect causes correlations to be initially 
(for $\gamma \Delta t<-3k_y/k_x$) 
enhanced by the shear and only successively dumped.

Integrating the expression of Eq.~(\ref{trans}) over the space 
of wave vectors  we obtain the behaviour of the autocorrelation function
\begin{equation}
A(\Delta t)=\int _{k<\Lambda}\frac{d\vec k}{(2\pi)^d}C(\vec k,\Delta t)=
\frac{T}{2^{\frac{3d}{2}}\pi^{\frac{d}{2}}
\left [D(\infty)\Delta t\right ]^{\frac{d}{2}}
\left [1+\frac{7}{16}(\Delta t)^2\gamma ^2\right ]}
\label{pal}
\end{equation}
For $\Delta t < \Delta t^*=4/(\sqrt{7}\gamma )$ shear is ineffective
and $A$ decays as for $\gamma =0$. At large strain the 
term proportional to $\gamma$ in Eq.~(\ref{pal}) prevails so that
$A\sim \gamma ^{-1}(\Delta t)^{-(1+d/2)}$. Correlations are then 
suppressed by the flow for $\Delta t > \Delta t^*$. 
A relaxation time proportional to $\gamma ^{-1}$ is also
found in molecular dynamics simulations of supercooled liquids in shear flow
\cite{12}.

In this paper we have studied the behavior of a quenched 
glassy system under the effect of an applied shear flow.
The analysis has been carried out analytically in the 
framework of the mean field
model recently introduced in \cite{1}.
Two shear regimes can be distinguished
which behave differently. 
While for weak shear the off-equilibrium evolution
is unaffected by the flow, for strong shear 
the evolution towards the stationary state is enhanced and 
aging is interrupted
after a time $\tau _g=1/(T\gamma)$ by the flow itself.  
We have also described how the properties of the time
correlation function in the stationary state are modified by the flow.
These predictions are amenable of experimental
checks; it would be interesting to know if similar feature can be observed
in real glassy systems such as hard spheres or colloidal suspensions.

\stars

F.C. is grateful to M.Cirillo, R. Del Sole and M.Palummo 
for hospitality in the University of Rome.
F.C. and G.G. acknowledge support by the TMR network contract ERBFMRXCT980183
and by PRA-HOP 1999 INFM and MURST(PRIN 99).

\begin{figure}
\includegraphics*[width=18 cm,height=18 cm]{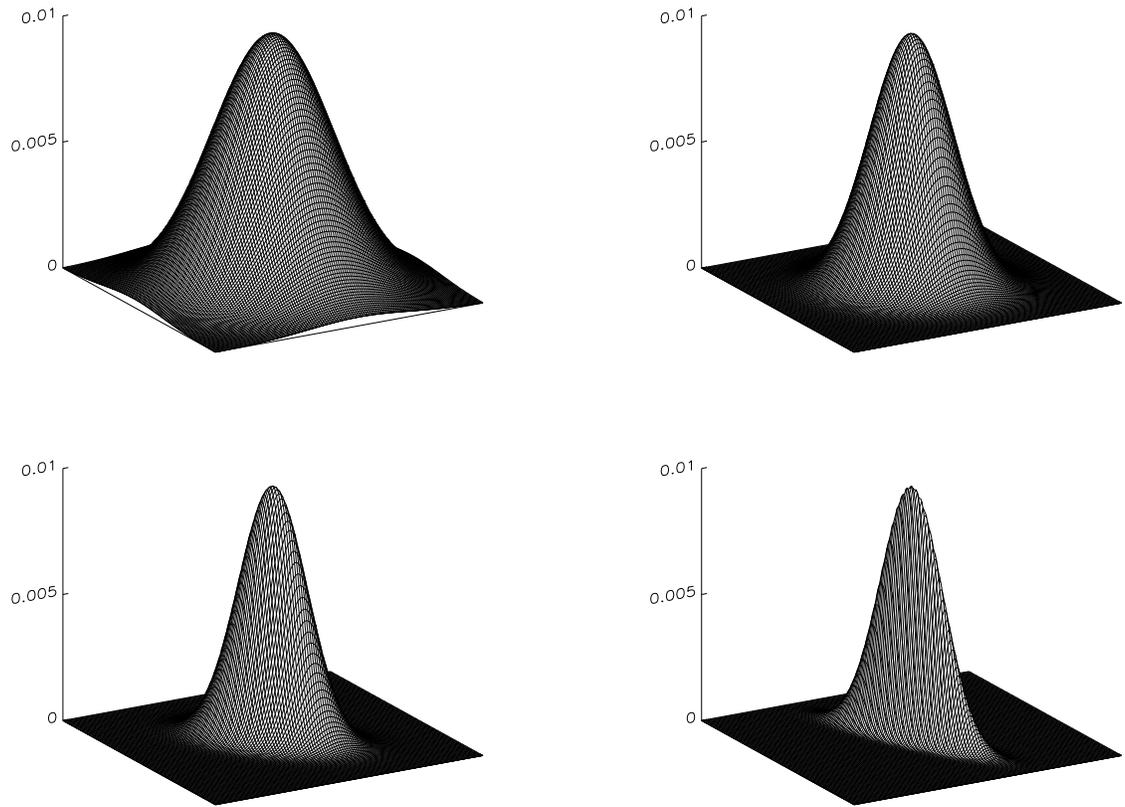}
\caption{Four snapshots of the structure factor $\chi (\vec k,t)$ 
are shown for $d=2$ at times $t= 2,7,20,60$ for $\rho =0.95$, $T=10^{-4}$ 
and $\gamma = 1$. $k_x$ increases from left to right ($k_x=0$ is in the
middle); $k_y$ increases towards the upper part of the figure.}
\end{figure}

\begin{figure}
\includegraphics*[width=15.0 cm,height=13.0 cm]{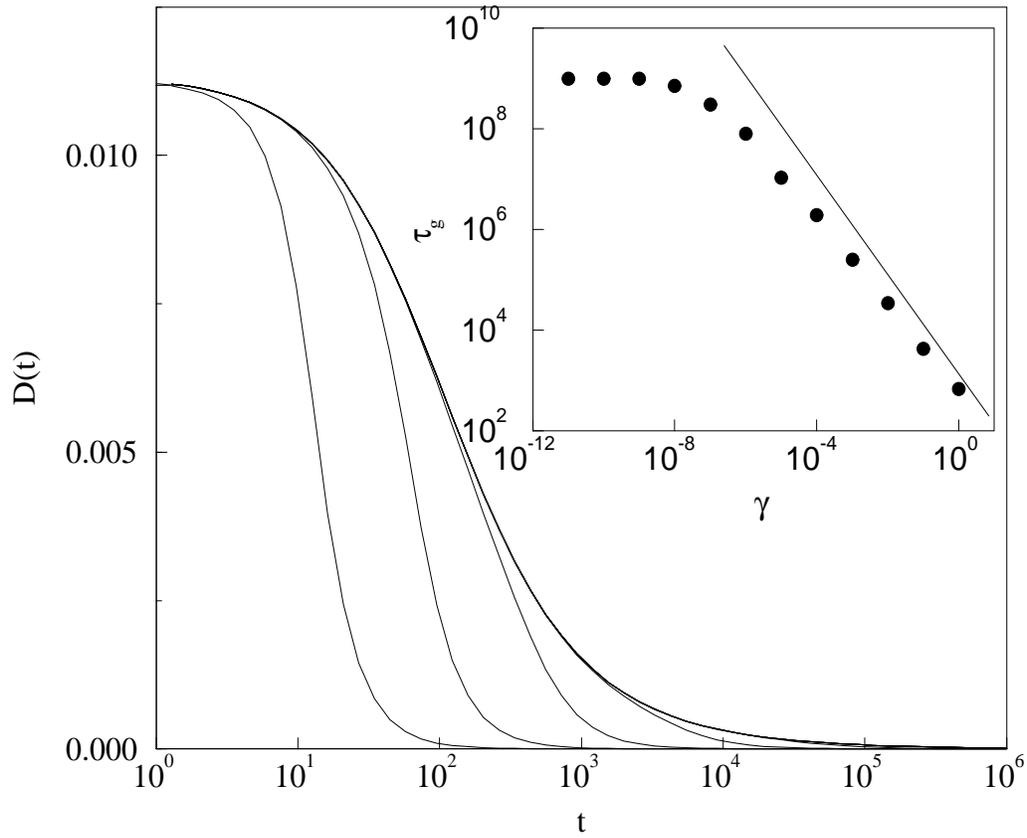}
\caption{$D(t)$ is shown in $d=2$ for $\rho = 0.95$, $T=10^{-4}$ 
and different values of $\gamma$
($\gamma = 1, 10^{-1}, 10^{-2}, 10^{-3}, 10^{-8}, 0$) from left to right).
The curves with $\gamma =10^{-8}$ and $\gamma =0$ are not
distinguishable in this plot. The finite final value $D(\infty)$
is very small for high densities and 
practically coincides with the $t$ axis in this picture.
In the inset the characteristic
time $\tau _g$ when aging is interrupted is plotted against $\gamma $.
The solid line is the $\gamma ^{-1}$ law.} 
\end{figure}

\begin{figure}
\includegraphics*[width=15.0 cm,height=13.0 cm]{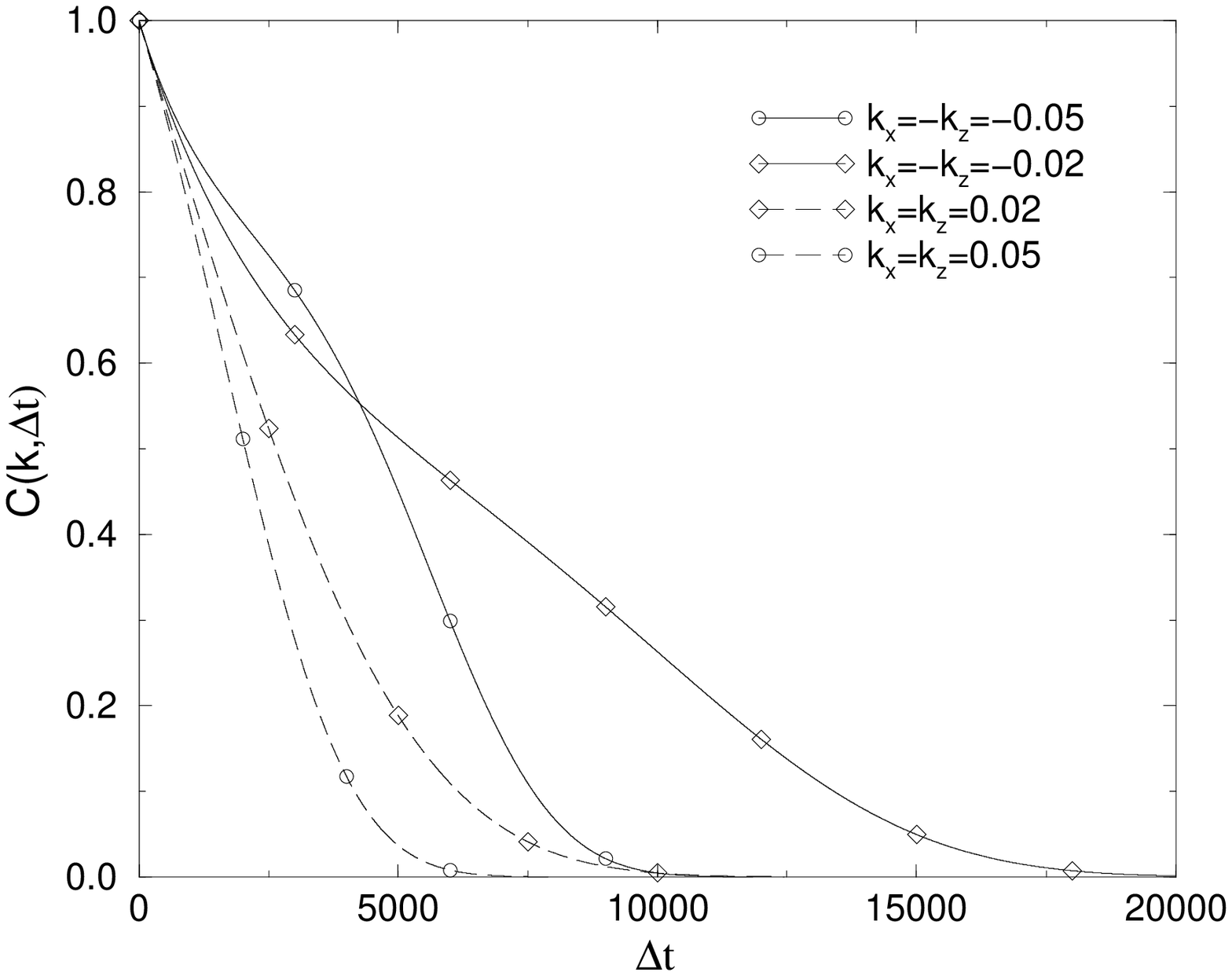}
\caption{The two-time correlator $C(\vec k,\Delta t)$ in the stationary
state is plotted against $\Delta t$ for $\gamma =10^{-2}$, $k_y=1$
and different values of $k_x$ and $k_z$ as indicated in the legend.
Solid lines refer to wavevectors in the sector with $k_xk_y<0$
whereas dashed lines correspond to the region $k_xk_y>0$.}
\end{figure}


\begin{thebibliography}{99}
\bibitem{11}
Equilibration can also be achieved by the {\it spontaneous} relaxation
of dense regions even at $T=0$. However it can be shown \cite{1}
that, in the limit $\overline \rho \to 1$, this process becomes so slow
that equilibration is always driven by thermal fluctuations.
\end{thebibliography}
\end{document}